\documentclass[usegraphicx]{mn2e}
\usepackage{psfig}
\renewcommand\[{\begin{equation}}
\renewcommand\]{\end{equation}}
\def\ex#1{\left\langle#1\right\rangle}
\def\kpc{\,{\rm kpc}}

\def\Myr{\,{\rm Myr}}\def\Gyr{\,{\rm Gyr}}
\def\pc{h^{-1}\,{\rm pc}}
\def\msun{h^{-1}\,{\rm M_{\odot}}}
\def\kms{\,{\rm km s}^{-1}}
\def\pa{\partial}

\def\d{{\rm d}}\def\e{{\rm e}}
\def\p{\partial}
\def\i{\relax\ifmmode{\rm i}\else\char16\fi}
\def\lesssim{{_ <\atop{^\sim}}}
\def\lta{\lesssim}
\def\grtsim{{_ >\atop{^\sim}}}
\def\gta{\grtsim}

\def\fracj#1#2{{\textstyle{#1\over#2}}}
\def\b#1{{\bf{#1}}}

\def\lesssim{\mathrel{\hbox{\rlap{\hbox{\lower4pt\hbox{$\sim$}}}\hbox{$<$}}}}
\def\gtrsim{\mathrel{\hbox{\rlap{\hbox{\lower4pt\hbox{$\sim$}}}\hbox{$>$}}}}

\def\apj#1 #2 {ApJ, {\bf #1}, #2}
\def\aj#1 #2 {AJ, {\bf #1}, #2}
\def\mn#1 #2 {MNRAS, {\bf #1}, #2}
\def\aa#1 #2 {A\&A, {\bf #1}, #2}
\def\dint{\int\!\!\!\!\int}
\def\gv{{\bf g}}
\def\gvz{\gv_0}

\def\gz{g_0}

\def\rhoo{\rho_1}
\def\rhos{\rho_*}
\def\rhod{\rho_{\rm DM}}

\def\phio{\Phi_1}

\def\gn{g_{\rm N}}

\def\ao{a_0}
\def\ro{r_0}

\def\rh{r_{\rm h}}
\def\msun{M_{\odot}}
\def\acc{cm s$^{-2}\;$}
\def\dvp{\Delta v_{\perp}}

\def\grad{\nabla}
\def\mone{m_1}
\def\mtwo{m_2}

\def\ms{m_*}
\def\md{m_{\rm DM}}
\def\mf{m_{\rm f}}
\def\ma{m_{\rm a}}

\def\avx{v_x}
\def\tb{t_{\rm 2b}}

\def\trh{t_{\rm rh}}

\def\tcross{t_{\rm cross}}
\def\tbM{\tb^{\rm M}}
\def\tbN{\tb^{\rm N}}

\def\tfric{t_{\rm fric}}
\def\tfricM{\tfric^{\rm M}}
\def\tfricN{\tfric^{\rm N}}

\def\xv{{\bf x}}

\def\vv{{\bf v}}
\def\DM{{\cal R}}
\def\vtyp{v_{\rm typ}}

\def\vi{v_i}
\def\vj{v_j}
\def\wSigma{\widetilde\Sigma}

\def\Ylm{Y^m_l}

\def\Yom{Y^m_1}
\def\Yomc{Y^{m*}_1}
\def\Yoz{Y^0_1}

\def\Ms{M_*}
\def\Mdm{M_{\rm DM}}
\def\Mtot{M_{\rm T}}

\begin{document}

   \title[Two-body relaxation in MOND]
   {Two-body relaxation in modified Newtonian dynamics}

   \author[Ciotti \& Binney]
          {Luca Ciotti$^1$ \& James Binney$^2$
           \\
		   $^1$Astronomy Department, University of Bologna, 
                       via Ranzani 1, 40127 Bologna, Italy\\
                   $^2$Oxford University, Theoretical Physics, 
                        1 Keble Road, Oxford OX1 3NP\\
          }

   \date{Accepted version}

   \maketitle

\begin{abstract} 
A naive extension to MOND of the standard computation of the two-body
relaxation time $\tb$ implies that $\tb$ is comparable to the crossing
time regardless of the number $N$ of stars in the system. This
computation is questionable in view of the non-linearity of MOND's
field equation. A non-standard approach to the calculation of $\tb$ is
developed that can be extended to MOND whenever discreteness noise
generates force fluctuations that are small compared to the mean-field
force. It is shown that this approach yields standard Newtonian
results for systems in which the mean density profile is either
plane-parallel or spherical. In the plane-parallel case we find that
in the deep-MOND regime $\tb$ scales with $N$ as in the Newtonian
case, but is shorter by the square of the factor by which MOND
enhances the gravitational force over its Newtonian value for the same
system. Near the centre of a spherical system that is in the deep-MOND
regime, we show that the fluctuating component of the gravitational
force is never small compared to the mean-field force; this conclusion
surprisingly even applies to systems with a density cusp that keeps
the mean-field force constant to arbitrarily small radius, and suggests that
a cuspy centre can never be in the deep MOND regime.
Application of these results to dwarf galaxies and groups and clusters
of galaxies reveals that in MOND luminosity segregation should be far
advanced in groups and clusters of galaxies, two body relaxation
should have substantially modified the density profiles of galaxy
groups, while objects with masses in excess of $\sim10\msun$ should
have spiralled to the centres of dwarf galaxies.

\end{abstract}

\section{Introduction}

Milgrom (1983) proposed that the failure of galactic rotation curves
to decline in Keplerian fashion outside the galaxies' luminous body
arises not because galaxies are embedded in massive dark haloes, but
because Newton's law of gravity has to be modified for fields that
generate accelerations smaller than some value $\ao$. Bekenstein \&
Milgrom (1984; hereafter BM84) proposed the non-relativistic field
equation, eq.~(\ref{BekMfield}) below, for the gravitational potential
$\Phi$ that generates an appropriately modified gravitational
acceleration $\gv$ of a {\it test\/} particle through 
\[\label{gfromPhi}
\gv=-\nabla\Phi.
\]
A considerable body of observational data now supports this theory of
modified Newtonian gravity (MOND) -- see Sanders \& McGaugh (2002) for
a review. The big problem with MOND is our inability to derive the
MOND field equation as the low-energy limit of a Lorentz covariant
theory. This inability is unfortunate in two respects. First it makes
it impossible to determine MOND's predictions for gravitational
lensing experiments, or any observation that involves relativistic
cosmology, particularly observations of the CMB.  Since these are the
areas in which the competing Cold Dark Matter (CDM) theory has been
most successful, the lack of a Lorentz covariant form of MOND makes a
fair confrontation between MOND and CDM impossible. Another reason to
regret this lack is that there are tantalizing hints that the
characteristic acceleration $\ao \simeq 2\times10^{-8}$ \acc that lies at
the heart of MOND is connected to the requirement for a non-zero
cosmological constant $\Lambda$ in Einstein's equations: $\ao$ and
$\Lambda\simeq 3(\ao/c)^2$ may be two aspects of a single physical
process associated with the unknown small-scale structure of
space-time (Milgrom 2002).

Clearly it is worthwhile to pursue these ideas regarding a putative
Lorentz covariant form of MOND only if the low-energy theory that we
already have accounts for all available data. In this paper we show
that MOND predicts two-body relaxation times for systems in which
$g\ll\ao$ that are shorter than those given by Newtonian theory by a
factor $\sim(g/\ao)^2$. We show further that in MOND dynamical
friction operates on a timescale that is shorter than in Newtonian
dynamics with dark matter by a factor $\sim g/\ao$.  The
shortness of the dynamical friction timescale has observationally
testable predictions for the dynamics of systems such as dwarf
galaxies, groups and clusters of galaxies that are in the deep MOND
regime.

\section{Order-of-magnitude analysis}

BM84 replace Poisson's equation for the potential in
terms of the density $\rho$ by
\[\label{BekMfield}
\grad\cdot [\mu (|\grad\Phi|/\ao)\grad\Phi ] = 4\pi G \rho,
\]
to be solved subject to the boundary condition $|\grad\phi|\to 0$ for
$|\xv |\to\infty$. The function $\mu(x)$ is required to have the
behaviour
\[\label{mulimits}
\mu\simeq\cases{x&for $x\ll1$,\cr1&for $x\gg1$, }
\] 
but the detailed manner in which $\mu$ moves between these limits
is currently constrained by neither observational data nor theory. 

We
shall be concerned with the `deep MOND regime' in which $\mu(x)\simeq
x$. In this limit equation (\ref{BekMfield}) can be written
 \[
\grad\cdot \left[{|\grad\Phi|\grad\Phi\over a_0}-\grad\Phi_{\rm N} \right] = 0,
\]
 where $\Phi_{\rm N}$ is the Newtonian potential generated by the given
density distribution. This equation implies that the difference between the
two terms in the square brackets is equal to the curl of some vector field.
BM84 show that when the density distribution is spherical, planar or
cylindrical, the curl vanishes. It then follows that the acceleration ${\bf
g}_0$ in MOND is related to the Newtonian acceleration ${\bf g}_{\rm N}$ by
 \[\label{deepM}
g_0^2=a_0g_{\rm N}.
\]

We now present a heuristic derivation of the two-body relaxation time
of a system that is in this regime. Our analysis is modelled on the
standard Newtonian derivation of the two-body time (e.g., \S4.1 of
Binney \& Tremaine 1987, hereafter BT).  This derivation is
straightforward to follow and illuminates the basic physical principle
that causes two-body relaxation to be fast in MOND.  It is open to the
criticism, however, that it ignores the inherent non-linearity of the
basic equation (\ref{BekMfield}).  Consequently, in the following
section we rederive the Newtonian relaxation time by a very different
technique that carries effortlessly over to the case of MOND.

Milgrom (1986; hereafter M86) shows that in the deep MOND regime
equation (\ref{BekMfield}) causes a force $F$ to act between two
isolated point masses $\mone$ and $\mtwo$, which can be written
\[\label{Feq}
F={G\mone\mtwo\over r^2}
f\left({\mtwo\over\mone},{r\over\ro}\right ),
\]
where
\[\label{defsro}
\ro\equiv \sqrt{G(\mone+\mtwo)\over\ao}
\simeq 8.1\, 10^{16}
\sqrt{{\mone+\mtwo\over\msun}}\quad {\rm cm}.
\]
The function $f$ in equation (\ref{Feq}) can be calculated numerically
as a function of $r/\ro$ for a given mass ratio. In particular,
numerical/asymptotic solution of equation (\ref{BekMfield}) shows
that:
\begin{itemize}
\item for $r/\ro \leq 1$, $f\sim 1$ independently of the mass ratio;

\item for $\mtwo\ll\mone$, the  acceleration  of
$\mtwo$ is given by equation (\ref{gfromPhi}), while the acceleration
of $\mone$ is obtained from the conservation of linear momentum
$\mone\gv_1+\mtwo\gv_2=0$;

\item finally, for nearly equal masses in the {\it deep-MOND regime},
\[\label{equalMeq}
F\simeq {\mone\mtwo\over\sqrt{\mone+\mtwo}}{\sqrt{G\ao}\over
r}= {G\mone\mtwo\over r\ro}.
\]

According to the numerical calculations of M86, this formula does not
err by more than 20\% for any mass ratio. Note that its asymptotic
behavior is correct when one of the masses is vanishingly small or much more
massive than the other.
\end{itemize}

For ordinary stars, $\ro$ is much less than the mean interstellar
distance in galaxies, so nearly all stellar interactions are in the
deep-MOND regime if the local gravitational mean field $g$ of the
whole galaxy is.  If, by contrast, $g\ga\ao$, then all interactions,
even ones that are individually weak, conform to Newtonian theory --
this result is sometimes called the ``external field effect''
(BM84). An immediate consequence of this result is that MOND predicts
standard two-body relaxation times for objects in which the mean field
exceeds $\ao$, even though individual stellar interactions generate
accelerations much smaller than $\ao$.

Consider then a stellar encounter in a system that is in the deep-MOND
regime. For simplicity we assume that both stars have the same mass
$m$.  Let the encounter be characterized by impact parameter $b\geq
\ro$ and asymptotic relative velocity ${\bf V}$.  Using equation
(\ref{equalMeq}) and the impulse approximation, we conclude that the
encounter changes the velocity of each star (in the direction
perpendicular to ${\bf V}$) by an amount
\[\label{givesdvp}
\dvp\simeq {2b\over V}\times {F(b)\over m} = {2 Gm\over\ro V}.
\]
Remarkably, this formula does not contain the impact parameter $b$:
{\it in the deep-MOND regime, the deflection is independent of the
impact parameter}. The steady accumulation of such velocity changes
causes the star's velocity to execute a random walk. If the system
contains $N$ stars and has half-mass radius $R$, per crossing time the
star experiences
\[
\delta n={N/2\over\pi R^2}2\pi b\d b
\]
encounters with impact parameter in $(b+\d b,b)$. Adding in quadrature
the velocity changes from successive encounters on the assumption that
they are uncorrelated yields a cumulative change in the square of the
stellar speed per crossing time
\[\label{givesDeltav} 
(\dvp^2)_{\rm cross}={N\over R^2}\left({2Gm\over\ro V}\right)^2 
               \int_0^R\d b \,b=2N\left({Gm\over\ro V}\right)^2.
\]
The ratio of the two-body time to the crossing time is thus
\[
{\tb\over\tcross}\simeq {\vtyp^2\over(\dvp^2)_{\rm cross}}=
                        {\vtyp^4\ro^2N\over 2G^2M^2},
\]
where $M=Nm$ is the system's mass and $\vtyp$ a typical stellar
velocity.  Finally we use equation (\ref{defsro}) and the fundamental
MOND equation $\vtyp^4=GM\ao$ to eliminate $\ro$ and $\vtyp$ from this
equation, and find
\[
\tb =\tcross.
\]
Thus this naive calculation implies that in the deep-MOND regime the
two-body relaxation time is comparable to the crossing time regardless
of the number of stars $N$. The physical origin of this result is
clear: equation (\ref{givesdvp}) states that in the deep-MOND regime
distant encounters produce much larger deflections than in the
Newtonian case, and even in the latter case distant encounters make a
large contribution to the relaxation rate.

\section{Relaxation in  a uniform field}

The derivation of $\tb$ that we gave in the last section is
objectionable on two grounds. First, in MOND all two-body orbits are
bound because the potential is asymptotically logarithmic. So our use
of the impulse approximation is highly suspect. Second, the derivation
assumes that the effects of encounters can simply be added. The
dominant encounters are those at impact parameters comparable to the
half-mass radius. Each such encounter lasts of order a crossing time,
so the many encounters that contribute to $(\dvp^2)_{\rm cross}$ in
equation (\ref{givesDeltav}) occur simultaneously. In the deep-MOND
regime the field equation (\ref{BekMfield}) is highly non-linear, and
it is far from clear that the effects of different encounters can be
simply added. In this section we derive $\tb$ by a different approach
that is not open to this objection.

The underlying physical idea is that two-body relaxation is driven by
fluctuations in the gravitational potential of a system that is in
virial equilibrium. The fluctuations are generated by Poisson
noise. We obtain $\tb$ by decomposing the fluctuations into different
spatial frequencies, and summing over frequencies rather than over
encounters.  We shall find that with this approach, the calculation of
$\tb$ for MOND differs very little from the corresponding Newtonian
calculation. Hence we now rederive the Newtonian relaxation rate with
the new formalism, to demonstrate that it produces the familiar
result, and to pave the way for the calculation of $\tb$ in MOND.

\subsection{Newtonian relaxation}

We wish to consider the case in which the underlying system generates
a uniform (Newtonian) gravitational field $\gv$. Such a field is
generated by an infinite sheet of constant density. So we consider the
effect that density fluctuations in this sheet have on a star that is
located distance $z$ from the sheet.  Let $\xv$ be a two-dimensional
vector of coordinates in the plane of the sheet. Then at $z\ne0$
Laplace's equation has solutions (e.g.\ BT \S5.3.1)
\[\label{Lapsoln}
\Phi(\xv)=\int\d^2\check\b k\,
\widetilde\Phi(\b k)\e^{-k|z|}\exp(\i\b k.\xv),
\]
where $\widetilde\Phi$ is an arbitrary function, $k=|\b k|$, and
$\d^2\check\b k\equiv\d^2\b k/(2\pi)^2$.  If the surface density of
the sheet is
\[
\Sigma(\xv)=\int\d^2\check\b k\,\wSigma(\b k)\exp(\i\b k.\xv),
\]
then an application of Gauss's theorem shows that
\[\label{SigPhi}
\widetilde\Phi(\b k)=-2\pi G{\wSigma(\b k)\over k}.
\]
Differentiating (\ref{Lapsoln}) with respect to $x$ and integrating
with respect to time, we calculate a component of velocity parallel to
the sheet that the fluctuating density induces in our test particle:
\begin{eqnarray}
\avx(\tau)&=&-\int_0^\tau\d t\,{\p\Phi\over\p x}\nonumber\\
          &=&2\pi G\int_0^\tau\d t\int\d^2\check\b k\,
{\i k_x\over k}\wSigma(\b k,t)\e^{-k|z|}\exp(\i\b k.\xv).
\end{eqnarray}
Squaring $\avx$ and taking an ensemble average, we have
\begin{eqnarray}\label{exwant}
\ex{\avx^2(\tau)}&=&-(2\pi G)^2\int\d^2\check\b k\int\d^2\check\b k'
                     \e^{-(k+k')|z|}{k_xk_x'\over kk'}\nonumber\\
&&\hskip-.6cm\times\int\d t\,\int\d t'\,\exp[\i(\b k +\b k').\xv]
\ex{\wSigma(\b k,t)\wSigma(\b k',t')}.
\end{eqnarray}
Let $\xv_\alpha$ denote the location within the sheet of star
$\alpha$.  Then
 \[
\Sigma(\xv,t)=m\sum_\alpha\delta[\xv_\alpha(t)-\xv]
\]
 so
\[
\wSigma(\b k,t)=m\sum_\alpha\exp[-\i\b k.\xv_\alpha(t)].
\]
For the moment we assume that the velocity $\vv_\alpha$ of star $\alpha$
lies within the sheet. Then to a sufficient
approximation $\xv_\alpha(t')=\xv_\alpha(t)+(t'-t)\b v_\alpha$ and
we have for uncorrelated stars
\begin{eqnarray}\label{exzero}
\ex{\wSigma(\b k,t)\wSigma(\b k',t')}&=&m^2\nonumber\\
&&\hskip-2cm\times\ex{\sum_\alpha\exp[-\i\xv_\alpha.(\b k+\b k')]
\exp[\i(t'-t)\b k'.\b v_\alpha]}.
\end{eqnarray}
Since the velocities of stars are independent of their positions, the
ensemble average above can be expressed as the product of two ensemble
averages, one over $\xv_\alpha$ and the other over $\b v_\alpha$.
With $\b K=\b k+\b k'$ we have
\[
\ex{\exp(-\i\xv_\alpha.\b K)}=\int_A{\d^2\xv\over A}
     \exp(-\i\xv.\b K)={(2\pi)^2\over A}\delta(\b K),
\]
where $A$ is any large area of the sheet, so
\[\label{exone}
\sum_\alpha\ex{\exp(-\i\xv_\alpha.\b K)}=(2\pi)^2n\delta(\b K),
\]
where $n$ is the number of stars per unit area. For the other ensemble
average we have for a Maxwellian distribution of velocities
\begin{eqnarray}\label{extwo}
\ex{\exp[\i(t'-t)\b k'.\b v_\alpha]}&=&\int{\d^2\b v\over2\pi\sigma^2}
\e^{-v^2/2\sigma^2}\exp[\i(t'-t)\b k'.\b v]\nonumber\\
&=&\exp\left[-{\textstyle{1\over2}}(t'-t)^2k^{\prime2}\sigma^2\right].
\end{eqnarray}
Substituting equations (\ref{exzero}), (\ref{exone}) and (\ref{extwo})
into equation (\ref{exwant}) and integrating over $t'$ and $t$, we
find for $\tau\gg k\sigma$
\begin{eqnarray}\label{sheetDC}
{\ex{\avx^2(\tau)}\over\tau}&=&{(2\pi)^{5/2}G^2m^2n\over\sigma}
\int\d^2\check\b k\,\e^{-2k|z|}{k_x^2\over k^3}\nonumber\\
                            &=&{(2\pi)^{3/2}G^2m^2n\over4|z|\sigma}.
\end{eqnarray}

From equation (\ref{sheetDC}) we can recover the standard expression
for the Newtonian diffusion coefficient by summing the contributions
from many sheets. If $\rho$ is the (homogeneous) mass density due to
stars, then $\rho\d z=nm$, and the overall diffusion coefficient
is
\[
{\ex{\avx^2(\tau)}_z\over \tau}=
{(2\pi)^{3/2}G^2\rho m\over2\sigma}\!\int_0^\infty\!{\d z\over z}=
{(2\pi)^{3/2}G^2\rho m\over2\sigma}\ln\Lambda ,
\]
with $\Lambda=z_{\rm max}/z_{\rm min}$. This diffusion coefficient may
be compared with half the value of $D(\dvp^2)$ in equation (8-68) of
BT. Taking the limit $X\to0$ of small test-particle velocities we find
\[
{2\ex{\avx^2(\tau)}_z\over\tau D(\dvp^2)}={6\pi\over 8}.
\]
 Thus this derivation agrees with the classical one to within the
uncertainties inherent in either approach. The weakest part of the present
derivation is the assumption that the velocities of field stars lie within
planes $z=\hbox{constant}$. Relaxing this assumption would reduce the
auto-correlation of each sheet's surface density below that given by
equation (\ref{exzero}) and introduce correlations between the densities of
different sheets.  When the change in $z$ during an encounter is small
compared to the distance of the particle from the point of observation,
little will have changed physically from the case of constant $z$, so the
new correlations will almost exactly compensate for the lost contribution to
the autocorrelation. This argument suggests that the error introduced by
confining particles to planes of constant $z$ is not large, as is also
indicated by the agreement between our value of the diffusion coefficient
and that obtained in  the standard way.

\subsection{Relaxation in MOND}

So long as the total number of stars in the system is large, and we
avoid special points of symmetry such as the system's centre, the
fluctuations in the gravitational field are small compared to the
MOND mean field $\gv$. Hence we may linearize the field equation
(\ref{BekMfield}) around $\gv$ and solve a linear field equation for
the component of the potential that drives relaxation.  As in the
Newtonian case, we decompose the perturbing density field into sheets
normal to the mean gravitational field. We find the potential
fluctuations due to each sheet, and add the effects of the sheets.

M86 shows that when the field equation (\ref{BekMfield}) is perturbed
about a uniform gravitational field $\gvz$, the first-order
perturbations to the potential and density are connected by
\[
\left(\nabla^2+{\p^2\over\p z^2}\right)\Phi={\ao\over\gz}4\pi G\rho,
\]
 where $\nabla^2$ is the full three-dimensional Laplacian operator.
Thus the solutions (\ref{Lapsoln}) to Laplace's equation must be replaced by
\[\label{MONDsoln}
\Phi(\xv)=\int\d^2\check\b k\,
\widetilde\Phi(\b k)\e^{-k|z|/\surd2}\exp(\i\b k.\xv),
\]
and equation (\ref{SigPhi}) becomes
\[\label{MONDSigPhi}
\widetilde\Phi(\b k)=-{\ao\over\surd2 \gz}2\pi 
                      G{\wSigma(\b k)\over k}.
\]
The Newtonian calculation carries over with two substitutions:
$|z|\to|z|/\surd2$ and $m\to (\ao/\surd2 \gz)m$. Equation
(\ref{sheetDC}) for the contribution of a single sheet to the
diffusion coefficient becomes
\[\label{diffCM}
{\ex{v_x^2(\tau)}\over\tau}={(2\pi)^{3/2}G^2m^2n\over 4|z|\sigma}
                            {\ao^2\over\surd2 \gz^2},
\]
so the diffusion coefficient is larger than in the Newtonian case by a
factor $\ao^2/(\surd2 \gz^2)$. 

Consider now two systems that contain identical distributions of stars
in phase space, but differ in the way gravity works: in one system
gravity is Newtonian, while in the other it is described by MOND. In
the Newtonian case we augment the gravitational field of the stars
with a fixed background field to ensure overall dynamical
equilibrium. Then the two-body relaxation times in the two systems are
in the inverse ratio of their diffusion coefficients, which from equations
(\ref{sheetDC}) and (\ref{diffCM}) is
\[\label{rat2b}
{\tbM\over\tbN}={\surd2\gz^2\over\ao^2}.
\]
This result will remain true when the fixed background field in the
Newtonian system is replaced by the field generated by a distribution
of DM particles provided individual DM particles are much lighter than
stars -- the particles then make negligible contributions to the
fluctuations in the overall gravitational acceleration\footnote{In a
Newtonian system with field particles of two different species, with
masses $\ms$ and $\md$ and densities $\rhos$ and $\rhod$, the two
body relaxation time of a test particle of mass $\ma$ is $\tbN\propto
1/(\ms\rhos +\md\rhod)$.}.

In the deep-MOND regime, equation (\ref{deepM}) holds, so eliminating $\ao$
from equation (\ref{rat2b}), the ratio of two-body times becomes
 \[\label{2bratio}
{\tbM\over\tbN}={\surd2\gn^2\over\gz^2}={\surd2\over(1+\DM)^2},
\]
where
\[
\DM\equiv{\Mdm\over\Ms}
\]
is the ratio of the DM to stellar mass in the Newtonian system.

\subsection{Dynamical friction}

Equation (\ref{2bratio}) has important implications for the magnitude
of dynamical friction in MOND because the dynamical friction
experienced by a body as it moves through a population of background
objects that are in thermal equilibrium, is proportional to its
diffusion coefficient in velocity space.  This relation can be
verified for coefficients calculated under the assumption of standard
Newtonian gravity, but, as Chandrasekhar (1943) pointed out, it
follows from the general principles of statistical mechanics. This
fact ensures that the relation is valid regardless of the law of
gravity.

In the local approximation, the Fokker-Planck equation for the
evolution of the distribution function $f$ of a population of `test'
objects can be written
\[
{\d f\over\d t}=-{\p\over\p\b v}\cdot\b S,
\]
 where the flux $\b S$ of stars in velocity space is [BT eq.~(8-57)]
\[
S_i=-fD_i+\fracj12\sum_j{\p\over\p \vj}(fD_{ij}).
\]
Here $D_i=D(\Delta\vi)$ and $D_{ij}=D(\Delta\vi\,\Delta\vj)$ are the
usual first- and second-order diffusion coefficients.  Let the
scattering objects have mass $\mf$ and be in thermal equilibrium at
inverse temperature $\beta=(\mf\sigma^2)^{-1}$. Then the principle of
detailed balance implies that $\b S$ will vanish when
$f\sim\exp(-\beta H)$, where $H=\ma(\fracj12v^2+\Phi)$ is the
Hamiltonian that governs the motion of the test objects.  In a frame in which
$D_{ij}$ is diagonal it then follows that
\begin{eqnarray}\label{friceq}
D_i&=&-\fracj12\beta\ma\vi D_{ii}+\fracj12{\p\over\p \vi}D_{ii}\nonumber\\
   &=&-\left({\ma\over 2\mf}-\sigma^2{\p\over\p \vi^2}\ln D_{ii}\right)
        {D_{ii}\over\sigma^2}\vi
\end{eqnarray}
(no sum over repeated indices).  For $\vi\lta\sigma$ the term
with the logarithmic derivative is of order unity -- in the case of a
Maxwellian distribution of scatterers it evaluates to $-3/10$ at
$\vi=0$ [BT, eq.~(8-65)].  Hence for $\ma\gta4\mf$ equation
(\ref{friceq}) implies that $\vi$ decays exponentially in a
characteristic time $\tfric$ that is related to the two-body
relaxation time by
\[\label{fricrat}
{\tfric\over\tb}={2\mf\over\ma};
\]
this result is consistent with the Spitzer (1987) relation
$\tfric/\tb=2\mf/(\mf +\ma)$, when $\ma\gta4\mf$. From this result and
eq.~(\ref{2bratio}) it follows that in MOND the friction time is
reduced by a factor $(1+\DM)^2/\surd2$ over the value it would have in
a Newtonian system with the same stellar mass and a {\it fixed\/}
auxiliary gravitational field.

Whereas the diffusion coefficients were unchanged when the fixed
gravitational field was replaced by the field of swarms of low-mass DM
particles, this replacement enhances dynamical friction by a factor
$(1+\DM)$. Hence\footnote{A system with comparable masses of luminous
and DM violates equation (\ref{fricrat}) because the stars and DM
particles are not in thermal equilibrium with one another.  The
standard treatment of dynamical friction shows that for a test
particle of mass $\ma$ $\tfricN\propto 1/[(\ma +\ms)\rhos
+(\ma+\md)\rhod]$, and from Footnote 1, when $\ma\gta$ of $\ms$ and
$\md$, and $\ms\gta\md$ equation (35) is reobtained.}.
\[\label{Nfricrat}
{\tfricN\over\tbN}={2\mf\over\ma}(1+\DM)^{-1},
\]
and 
\[
{\tfricM\over\tfricN}={\surd2\over1+\DM}.
\]

\section{Extension to spherical systems}

Given that stellar systems are frequently approximately spherical, and
rarely have the plane-parallel symmetry that we assumed in the last
section, we try to adapt the preceding calculation to a spherical
system.  The appendix derives an expression for the Newtonian
diffusion coefficient experienced by a star that is stationary near
the centre of a spherical system.

If we are to use perturbation theory to carry this Newtonian analysis
over to the deep-MOND regime, the unperturbed acceleration $\gz$
should be non-vanishing and less than $a_0$ at the location of the
test star. For simplicity the system's density profile is taken to be
a power law in radius $\rho\sim r^{-\gamma}$, and the requirement that
at all radii $\gz/\ao$ be of order but less than unity then implies
$\gamma=1$. We show that in this case the acceleration caused by the
fluctuations becomes comparable to $\gz$ as one approaches the centre,
and the linearized field equation ceases to be valid.  We infer from
this result that in the deep-MOND regime the centres of all systems
must be homogeneous, since non-linear fluctuations in the acceleration
will soon disrupt the cusp in the density profile that generates the
assumed constant acceleration $\gz$.  Unfortunately, perturbation
theory cannot be used to calculate the central relaxation time of a
system with a constant-density core.

Thus this analysis does not lead to a value of $\tb$ for systems in
the deep-MOND regime, but it does suggest that at the centres of these
systems two-body relaxation is very much more rapid than in the
Newtonian case.

\section{Astrophysical applications}

We now apply the results of Section 3 to stellar systems that in
Newtonian dynamics would turn out to be DM dominated, namely to
dwarf galaxies, to galaxy clusters and to galaxy groups: all these systems
are in the deep-MOND regime.

Equation (8-71) of BT gives the Newtonian two-body relaxation time for
a system with a given velocity dispersion $\sigma$ and density of
scatterers $\rhos$.  As in BT, we substitute into this equation
values of $\sigma$ and $\rhos$ appropriate to the system's half-mass
radius.  We obtain a result for the reference relaxation time that is
analogous to equation (8-72) of BT but different because now
$\sigma^2\simeq 0.4G\Ms (1+\DM)/\rh$ on account of the presence of
DM. With $\Lambda=0.4N(1+\DM)$ we have
\[
\trh^{\rm N} \simeq 0.66\Gyr
			{(1+\DM)^{3/2}\over \ln\Lambda} {\msun\over
			m}\left ({\rh\over {\rm pc}}\right )^{3/2}
			\left ({\Ms\over 10^5\msun}\right
			)^{1/2}\hskip-1em,
\]
while equation (\ref{2bratio}) now implies that the  relaxation
time for MOND is
\[
\trh^{\rm M} \simeq 0.9\Gyr
		  {(1+\DM)^{-1/2}\over\ln\Lambda}
          {\msun\over m}\left ({\rh\over {\rm pc}}\right )^{3/2}
          \left ({\Ms\over 10^5\msun}\right )^{1/2}\hskip-1em.
\]

\subsection{Dwarf galaxies}

For a dwarf galaxy such as Draco, $\Ms\simeq 2.6\,10^5\msun$,
$\rh\simeq 200\pc$, and $\DM\simeq 100$ (see. e.g., Mateo 1998, Kleyna
et al. 2002), so the reduction factor is enormous ($\sim
10^4$). However, $\trh^{\rm N}\simeq 10^{5}\Gyr$, so $\trh^{\rm M}$ is
still is slightly longer than the Hubble time.

Since in MOND the dynamical-friction time for inspiralling of an object of
mass $\ma$ is shorter than the two-body time by $\sim \msun/\ma$, any object
significantly more massive than a star will have spiralled to the centre of
a dwarf galaxy. In particular, any black holes with masses $\gta10\msun$
should have collected at the centre.  Since globular clusters have masses
$\gta10^4\msun$, they are liable to spiral to the galaxy centre even in
Newtonian dynamics. In MOND they spiral in on essentially a dynamical time.
Consequently, the possession of globular clusters by a dwarf galaxy that is
in the deep-MOND regime would be problematic for MOND. Interestingly, the
dwarfs with large values of $\DM$, namely Draco, UMi, Carina, and
Sextans, have no globular clusters listed in Table 8 of Mateo (1998).

\subsection{Galaxy clusters}
 
If $\Mtot$ is the total cluster mass, we have $\Ms\simeq0.05\Mtot$,
$M_{\rm gas}\simeq0.15\Mtot$, $\Mdm\simeq0.8\Mtot$ (see, e.g.,
B\"ohringer 1996), so $\DM\simeq 4$.  Adopting $N\simeq 100$,
$\Ms\simeq 1.7\, 10^{13}\msun$, $\rh\simeq 580\kpc$, and $\sigma
=1000\kms$, we find $\trh^{\rm N}\simeq 48\Gyr$ and $\trh^{\rm
M}\simeq2.7\Gyr$.

In MOND dynamical friction will cause a galaxy of mass $m_a$ to spiral
to the cluster centre on a timescale $\sim 5(\overline{m}/m_a)\Gyr$,
where $\overline{m}$ is the mass-weighted mean galactic mass. It
follows that in MOND luminosity segregation should be well advanced
within galaxy clusters.

\subsection{Galaxy groups}

For a typical galaxy group, $N\simeq5$, $\Ms\simeq 5.2\,
10^{11}\msun$, $\DM =16$, $\sigma\simeq 200\kms$, and
$\rh\simeq380\kpc$. Hence $\trh^{\rm N}\simeq 70\Gyr$ and $\trh^{\rm
M}\simeq 340\Myr$.  If one (improperly!) applied the concepts of
Newtonian stellar dynamics, one would conclude that in MOND galaxy
groups should have already evaporated because in the Newtonian case
the evaporation time is $\sim10\trh$. The logarithmic asymptotic form
of MOND's two-body potential implies that it is impossible to escape
from a perfectly isolated MOND system\footnote{We note that it is
generally supposed that the galaxy disks are made of shattered star
clusters and associations: while tidal destruction will still work,
the failure of evaporation in outlying parts of galaxies might lead to
the survival of surprising numbers of low-mass star
clusters.}. However, the shortness of $\trh^{\rm M}$ for galaxy groups
suggests that relaxation will have significantly increased the central
densities of groups. Luminosity segregation will proceed on a
dynamical timescale.

\section{Discussion and conclusion}

A naive extension to MOND of the standard derivation of the two-body
relaxation time $\tb$ leads to the conclusion that $\tb$ is comparable
to the crossing time for any value of the number $N$ of stars in the
system.  The derivation is open to objection on at least two grounds,
so we have developed an approach to the calculation of the Newtonian
relaxation time that can be straightforwardly adapted to MOND. This
focuses on the effect of density fluctuations, which must be small in
the limit of large $N$, and calculates the corresponding potential
fluctuations by linearizing MOND's field equation around a uniform
background field $\gz\ll\ao$. This analysis reveals that in MOND $\tb$
scales with $N$ in the same way as it does in the Newtonian case, but
is smaller by a factor $\sim(1+\DM)^2$, where $\DM$ is the ratio of
the apparent DM and stellar masses.

We show that the timescale $\tfric$ on which dynamical friction causes an
object of mass $\ma$ to spiral in through a population of scatterers of mass
$\mf$ is $\tfricM= (2\mf/\ma)\tbM$ in the case of MOND, while the Newtonian
time is $\tfricN\simeq (1+\DM)\tfricM$, provided individual DM particles
have negligible masses.

Application of these results to DM-dominated systems shows that at the
half-mass radius, the two-body relaxation times for MOND of both dwarf
galaxies and clusters of galaxies are of order the Hubble time.
Consequently, MOND predicts that objects in these systems that have
masses more than $\sim10$ times the mean mass will have spiralled to
the systems' centres. In particular, any black holes formed in
core-collapse supernovae should have collected at the centres of dwarf
galaxies, and luminosity segregation should be far advanced in
clusters of galaxies. The globular clusters of dwarf galaxies should
spiral inwards on a dynamical timescale, so the existence of globular
clusters in DM-dominated  dwarf galaxies would be problematic for MOND.

In MOND the two-body and dynamical-friction timescales of groups of
galaxies are only $\approx 300\Myr$. A Newtonian system with such a
short relaxation time would have evaporated completely, but
evaporation is probably impossible in MOND. However, in MOND
relaxation will surely have significantly modified the density
profiles of galaxy groups, and induced substantial luminosity
segregation. Further work is required to determine whether such
evolution is compatible with observation.

 The modification of Poisson equation given by equation (2) is only one
possible formulation fo MOND. We have not considered relaxation in modified
inertia theories, which give very similar predictions for rotation curves
and global mass discrepancies (Milgrom 2002). These theories could yield
very different predictions  for relaxation phenomena.

\section*{Acknowledgments}

We thank M. Milgrom for useful comments on the
manuscript.  L.C. would like to thank G. Bertin, J.P. Ostriker, and
T.S. van Albada for very useful discussions.

\appendix
\section{Relaxation at the centre of a spherical system}

\subsection{Newtonian case}

We consider the stochastic acceleration of a particle that is
initially stationary at the centre of a spherical system. We expand
the system's gravitational potential, generated by the (random)
density field
\[
\rhoo = \sum_{\alpha} m_{\alpha}\delta (\xv - \xv_{\alpha}),
\]
 in spherical harmonics. At the origin, only the dipole  term makes a
non-vanishing contribution to the force on a particle. Moreover, near the
origin the force contributed by the $l$th multipole varies as $r^{l-1}$
times a factor that changes sign each time the particle passes the origin
for even $l$. Hence averaged along a trajectory through the orgin, the force
from multipoles with $l>1$ is smaller by that from the dipole term by a
factor that vanishes with the square of the apocentric radius.
We therefore concentrate on the dipole term, which is
(BT \S2.4)
 \[
\label{basicphi}
\phio (r,\theta,\phi)=-\fracj43\pi G\sum_m\Yom (\theta,\phi)r
                             \int_r^\infty\d a\,\rho_{1m}(a),
\]
where
\begin{eqnarray}
\label{rhosum}
\rho_{1m}(a)&=&\int_{4\pi}\d^2\Omega\,\Yomc\rho(a,\theta,\phi)\nonumber\\
            &=&{1\over a^2}{\d\over\d a}\int_0^a\d r\,r^2
               \int_{4\pi}\d^2\Omega\,\Yomc\rho(r,\theta,\phi)\nonumber\\
            &=&{1\over a^2}{\d\over\d a}\int\d^3\b x\,\Yomc
               \sum_\alpha m_\alpha\delta(\b x-\b x_\alpha)\nonumber\\
            &=&{1\over a^2}{\d\over\d a}\sum_\alpha m_\alpha\Yomc (\alpha),
\end{eqnarray}
where the sum over $\alpha$ is extended to particles inside the sphere
of radius $a$. Differentiating (\ref{basicphi}) with respect a generic
direction (which without loss of generality we assume to be $z$), and
evaluating the obtained expression at $r=0$, we have
\[
\dot v_z=-{\pa\phio\over\pa z}=\fracj43\pi G\int_0^{\infty}
                                      \d a\,\rho_{10}(a)
\]
because the contributions from $m=\pm1$ vanish.  Integration with
respect to time gives
\[
v_z(\tau)=\fracj43\pi G\int_0^\tau\d t\,\int_0^{\infty}\d a\,\rho_{10}(a).
\]
Squaring and taking the ensemble average we find
\[
\label{givesexv}
\ex{v_z^2}={(4\pi G)^2\over9}\!\dint\! \d a\d a'\!\dint \d t\d t'
\,\ex{\rho_{10}(a,t)\rho_{10}(a',t')}.
\]
The expectation value in the integrand is non-negligible only for
$|a-a'|$ and $|t-t'|$ sufficiently small. With (\ref{rhosum}) we have
in the case that all particles have equal masses
\begin{eqnarray} 
C(a,t,t')&\equiv&\int\d a'\,\ex{\rho_{10}(a,t)\rho_{10}(a',t')}\nonumber\\ 
         &=&{m^2\over a^2}{\d\over\d a}\int
            {1\over a^{\prime2}}{\d\over\d a^{\prime}}
            \sum_{\alpha ,\beta}\ex{\Yoz (\alpha,t)\Yoz (\beta,t')}\nonumber\\
         &=& {m^2\over a^2}{\d\over\d a}\int
            {1\over a^{\prime2}}{\d\over\d a^{\prime}}
            \sum_{\alpha}\ex{\Yoz(\alpha,t)\Yoz(\alpha, t')},
\end{eqnarray}
where we have assumed that the particles are mutually uncorrelated. If
$\sigma$ is the characteristic tangential velocity of particle
$\alpha$, the expectation value in this equation vanishes after a time
of order $r_\alpha/\sigma$. Therefore we replace the integral over
$t'$ of $C(a,t,t')$ by $a/\sigma$ times half its peak value. To
estimate the latter we observe that by the Monte-Carlo theorem
\[
1=\int\d^2\Omega\,|Y_1^0|^2={4\pi\over N}\sum_{\alpha=1}^N|Y_1^0(\alpha)|^2.
\]
Hence
\[
\int\d t'\,C(a,t,t')={m^2\over8\pi a^2}{\d\over\d a}\left({1\over
a^2}{a\over\sigma}N\right).
\]
 Inserting this expression into (\ref{givesexv}) we have
\[
D\equiv
{\ex{v_z^2(\tau)}\over\tau}={2\pi (Gm)^2\over9\sigma}
\int{\d(N/a)\over a^2}.
\]
If the stellar density $\rho/m$ is a constant from some smallest
radius out to some maximum radius $r_{\rm max}$ and then zero, then
$N=\frac43\pi a^3(\rho/m)$ and the diffusion coefficient $D$ becomes
\[
D={16\pi^2 G^2m\rho\over27\sigma}
\ln\left({r_{\rm max}\over r_{\rm min}}\right).
\]
This should be compared with half the value of $\lim_{X\to0}D(\dvp^2)$
in eq. (8-68) of BT. One finds
\[
{2D\over\lim_{X\to 0}D(\Delta v^2_\perp)}={(2\pi)^{3/2}\over9}=1.75\ .
\]
As in the case of plane-parallel geometry we obtain a diffusion coefficient
that is larger by a factor $\la2$ than its conventional value.

\subsection{Case of the deep-MOND regime}

For the reasons given in the main text we assume that the system's
density profile obeys the power law $\rho\sim1/r$, which yields a
constant radial acceleration $\b\gz =-\sqrt{\ao GM_0/\ro^2}\,\b e_r$,
where $M_0$ is the mass interior to the fiducial radius $\ro$.

The perturbed potential $\Phi$ satisfies eq.\ (7) of M86 with $\mu_0=\gz/\ao$
and $L_0=1$ because we are in the deep-MOND regime. Hence
\[
{2\over r^2}{\p\over\p r}\left(r^2{\p\Phi\over\p r}\right)-
            {{\cal L}^2\over r^2}\Phi={\gz\over\ao}4\pi G\rho
\]
where ${\cal L}$ is the angular part of the Laplacian.  We seek the
dipole term in the expansion of $\Phi$ in spherical harmonics
\[
\Phi (r,\vartheta,\varphi) =\sum_{l=0}^{\infty}\sum_{m=-l}^{l}\Phi_{lm}(r)
\Ylm (\vartheta ,\varphi),
\]
which satisfies
\[
{1\over r^2}{\p\over\p r}
\left(r^2{\p\Phi_{10}\over\p r}\right)-{\Phi_{10}\over r^2}=
{\gz\over\ao}2\pi G\rho_{10}.
\]
The Green's function $u(r,r)$ for this equation satisfies
\[\label{Greeneq}
{1\over r^2}{\p\over\p r}\left(r^2{\p u\over\p r}\right)-
{u\over r^2}={\gz\over\ao}2\pi G\delta(r,r').
\]
 From the power-law solutions to the homogeneous equation it follows that
\[
u(r,r')=\cases{A(r')r^{\alpha-1/2}& for $r<r'$\cr
               B(r')r^{-\alpha-1/2}&for $r>r'$,}
\]
where $A(r)$ and $B(r)$ are functions to be determined from the rhs of
eq.\ (\ref{Greeneq}) and $\alpha=\sqrt{5}/2$. Consequently, near the
origin the dipole term varies as $r^{(\surd5-1)/2}\sim r^{0.62}$ and
the acceleration to which it gives rise diverges as $r^{-0.38}$. Hence
no matter how small the fluctuations in $\rho$ are, the linearization
of (\ref{BekMfield}) breaks down sufficiently near the origin. We
conclude that sufficiently near the origin the fluctuating part of the
field is comparable to, or larger than, $g_0$. In these circumstances
the fluctuations will soon disrupt the cusp in the density
distribution that generates $\gz$. It seems therefore that in the
deep-MOND regime the centres of all systems must be homogeneous.
Unfortunately, perturbation theory cannot be used to calculate the
central relaxation time of such a system. 

\end{document}